\documentclass[twocolumn,showpacs,superscriptaddress,longbibliography,nofootinbib,pra]{revtex4-1}
\usepackage[utf8]{inputenc}
\usepackage{calrsfs}
\usepackage{graphicx}
\usepackage{textcomp}
\usepackage{soul}
\usepackage{amsmath,amssymb}
\usepackage{color}
\usepackage{braket}
\usepackage{xcolor}
\usepackage{float}
\usepackage{physics}
\usepackage{soul}
\usepackage{graphicx}
\usepackage{caption}
\usepackage{amsmath}
\usepackage{amssymb}
\usepackage{bbold}
\usepackage{makecell}
\usepackage{tikz,pgfplots}
\usetikzlibrary{decorations.pathmorphing, arrows.meta, positioning, calc, shadings, shapes.geometric, shadows.blur, fadings}

\definecolor{gainRed}{RGB}{215, 48, 39}
\definecolor{lossBlue}{RGB}{69, 117, 180}
\definecolor{deepSpace}{RGB}{40, 40, 50}
\definecolor{gridLine}{RGB}{200, 200, 220}
\usepackage{mathtools}
\usepackage{multirow}
\usepackage{subcaption}
\usepackage{url}
\usepackage[colorlinks=true, urlcolor=blue, citecolor=blue, linkcolor=blue]{hyperref}
\begin{document}
\title{Traversability dynamics of minimal Sachdev-Ye-Kitaev Wormhole-inspired teleportation protocol with a parity-time ($\mathcal{PT}$)-symmetric non-Hermitian deformation}
\author{Sudhanva Joshi }
\email[]{sudhanvajoshi.rs.phy24@itbhu.ac.in}
\affiliation{Department of Physics, Indian Institute of Technology (Banaras Hindu University), Varanasi - 221005, India}

\author{Sunil Kumar Mishra}
\email[]{sunilkm.app@iitbhu.ac.in}
\affiliation{Department of Physics, Indian Institute of Technology (Banaras Hindu University), Varanasi - 221005, India}

\begin{abstract}
Holography-inspired teleportation has recently emerged as a significant area of research in quantum many-body systems. In this work, we investigate the effects of $\mathcal{PT}$ symmetric non-unitary deformations on the traversability of the wormhole-inspired teleportation protocol modeled by coupled Sachdev-Ye-Kitaev systems prepared in a Thermofield Double state bath. By introducing balanced gain and loss terms to the boundary Hamiltonians, we identify a phase transition driven by spectral exceptional points, where the real energy eigenvalues of the effective Hamiltonian coalesce and bifurcate into complex conjugate pairs. We demonstrate that the $\mathcal{PT}$-broken phase acts as an amplifier, enabling exponential growth in the norm of the teleported signal while preserving the causal time window for the wormhole's traversability. A statistical study of disorder realizations reveals that the critical non-Hermiticity threshold $\gamma_c$ follows a log-normal distribution, reflecting the sensitivity of the transition to the microscopic level spacing of the chaotic SYK spectrum. Furthermore, we observe a ``Purification" effect deep in the broken phase, where the teleportation channel acts as an entanglement distiller, yielding near-perfect teleportation fidelity for post-selected states. Our results suggest that the non-Hermitian topology can be harnessed to enhance holographic quantum communication, providing a robust mechanism for signal amplification in noisy, minimal quantum many-body systems.

\end{abstract}

\maketitle
\section{Introduction}

The Anti-de Sitter/Conformal Field Theory (AdS/CFT) correspondence is a conjectured duality equating a (d+1)-dimensional asymptotically Anti-de Sitter spacetime gravitational theory with a conformal field theory living on the (d)-dimensional boundary \cite{maldacena1999large, gubser1998gauge, klebanov1999ads}. This correspondence provides a concrete realization of the holographic principle by asserting that bulk gravitational dynamics are fully encoded in non-gravitational boundary quantum field theory \cite{bousso2002holographic}. Regimes of strongly coupled physics in the boundary theory, which maps to semi-classical gravitational regimes in the bulk, provide a powerful non-perturbative tool for studying quantum many-body systems which otherwise would be analytically inaccessible \cite{klebanov2001tasi, hubeny2015ads}. Over the past few years, the AdS/CFT correspondence has evolved significantly beyond its original string-theoretic context \cite{polyakov1998string} and has become a primary theoretical framework for exploring strongly correlated quantum systems, hydrodynamics, and quantum information \cite{sachdev2010strange,policastro2002ads,chen2022quantum}. In particular, the duality has offered geometric insights into entanglement, chaos, and thermalization, which are central concepts to modern many-body physics \cite{zizzi2000holography}. A notable development is the realization that certain quantum-mechanical models can exhibit emergent low-energy dynamics similar to those of near-$AdS_2$ gravity, thus providing solvable toy models for holography in the simplest possible setting. \\ 

The frontrunner for this type of paradigm is the Sachdev-Ye-Kitaev (SYK) model, which consists of $N$ Majorana fermions with random all-to-all couplings \cite{sachdev1993gapless,maldacena2016remarks}. Despite its conceptual simplicity, the SYK model exhibits a rich array of characteristic features, including maximal quantum chaos, non-Fermi liquid behavior, and a conformal regime at low energies, which suggests that its infrared dynamics are governed by an emergent reparametrization symmetry whose breaking is described by the Schwarzian action \cite{fu2017supersymmetric}. This action is the same action that governs fluctuations around $ AdS_2$ gravity, making the SYK model a leading microscopic realization of $AdS_2/CFT_1$ duality, offering tabletop access for investigating blackhole thermodynamics, scrambling dynamics, entanglement growth, and quantum gravitational phenomena within a solvable many-body system \cite{tezuka2023binary, polchinski2016spectrum,zhang2022quantum}. 
Realization of the whole ``Quantum gravity in Lab" paradigm has entailed the use of the SYK model and its variants to carry out wormhole-inspired teleportation \cite{brown2023quantum,nezami2023quantum,jafferis2022traversable,schuster2022many,Shapoval2023towardsquantum}. Wormhole-inspired teleportation protocol elucidates how traversable wormholes are realized via a double trace deformation that allows the information to pass through the bulk \cite{gao2017traversable}. In a traversable wormhole realized by the SYK Hamiltonian, an operator inserted on the left boundary can travel to the right boundary, during which its size does not monotonically grow; instead, its size distribution moves in a periodic way \cite{gao2021traversable}. When this size distribution obtains a complex phase, which is a linear function of the coupling parameter enables a coherent transmission of information. This property is referred to as perfect size winding \cite{nezami2023quantum}. \\
Currently established Wormhole-inspired teleportation protocols focus on unitary hermitian Hamiltonian evolution through the bulk, forming a closed system. However, realistic quantum systems are open to their environment, which naturally leads to non-Hermitian Hamiltonians with gain and loss terms \cite{cornelius2022spectral}. Here, a critical concept of $\mathcal{PT}$ symmetry comes into play. $\mathcal{PT}$-symmetric non-Hermitian Hamiltonians are a fascinating class of quantum systems that challenge traditional assumptions about quantum mechanics while still producing real energy spectra under certain conditions \cite{bender1998real}. In optics and photonics, there have been experimental realizations of $\mathcal{PT}$ symmetry in which the Schrodinger equation maps to the paraxial wave equation, in which there is a complex refractive index $n(x) = n_{R}(x) + \iota n_I(x)$ with real part $n_R$ corresponding to potential and imaginary part $n_I$ representing gain and loss \cite{ozdemir2019parity, feng2017non}. 
What happens when we introduce non-Hermitian topology (Exceptional Points) \cite{miri2019exceptional,heiss2012physics} into a holography-inspired system? Can gain and loss be used to enhance or control the traversability of a wormhole? \\
To find answers to these intriguing questions, we propose a $\mathcal{PT}$ symmetric extension of the wormhole-inspired teleportation protocol, introducing balanced gain and loss at the boundary. We introduce a non-Hermitian deformation via the operator $\rm H_{\mathcal{PT}} = i\gamma(Z_L-Z_R)$ where the $\gamma$ represents the strength of the drive and $Z_L$ and $Z_R$ are Pauli strings acting on the Left and Right boundary. Physically, $\gamma$ corresponds to a balanced injection of Gain into the input boundary and dissipation or loss at the output boundary. This setup enables us to investigate the interplay between the Hermitian scrambling dynamics responsible for spacetime emergence and the non-Hermitian topology induced by open system dynamics. \\
We then identify the microscopic origin of the phase transition in traversability as a spectral topological phenomenon. Here, spectral topology refers to the connectivity and braiding structure of the energy bands in the complex plane, governed by the presence of Exceptional Points. Through exact diagonalization, we show that the real energy eigenvalues of the effective Hamiltonian exhibit level attraction and coalescence as the non-Hermiticity $\gamma$ is increased. At a critical threshold $\gamma_c$, these levels undergo a bifurcation, emerging as complex conjugate pairs. We identify this transition as a high-order Exceptional Point (EP) in the many-body spectrum. This links holographic traversability to non-Hermitian spectral topology, demonstrating that the wormhole's stability is topologically protected in the exact $\mathcal{PT}$-symmetric phase and dynamically unstable in the broken phase. \\
We then move on to demonstrate that the $\mathcal{PT}$ broken phase functions as a ``Causal Amplifier" for quantum information. By analyzing the temporal phase diagram of the teleportation fidelity, we find that the non-Hermitian term generates exponential growth in the norm of the teleported signal, effectively acting as a gain medium for the traversing qubit. Crucially, our results show that this amplification respects the causal structure of the bulk geometry, as the scrambling time $t$, which defines the lightcone for traversability, remains invariant under the deformation. This suggests that non-Hermiticity can enhance signal recoverability without violating the causal connectivity of the emergent spacetime. \\
Finally, we characterize the protocol's robustness in the presence of disorder. Using the Random-matrix theory formalism, we reveal that the critical breakdown threshold $\gamma_c$ follows a log-normal distribution reflecting the hypersensitivity of the transition to the microscopic level spacing of the chaotic SYK Hamiltonian. Furthermore, we report a counterintuitive ``purification" effect deep in the broken phase ($\gamma \gg \gamma_c$). In this regime, the non-Hermitian evolution acts as a spectral filter, suppressing thermal noise components (loss modes) and projecting the system onto the optimal teleportation subspace (gain mode). This results in a fidelity plateau near unity for post-selected states, effectively distilling the entanglement and rendering the protocol robust against detuning in the geometric coupling strength. We then present our conclusion.

\section{Theoretical Framework}
\subsection*{SYK-Based Wormhole teleportation}
\begin{figure}
\centering

\begin{tikzpicture}[scale=0.85, xshift=-1.2cm]

% Background shading: Left and Right boundaries
\fill[cyan!10] (-4,3) .. controls (-2,2) and (-2,-2) .. (-4,-3) -- (-3.2,-3) .. controls (-1.8,-1.5) and (-1.8,1.5) .. (-3.2,3) -- cycle;
\fill[purple!10] (4,3) .. controls (2,2) and (2,-2) .. (4,-3) -- (3.2,-3) .. controls (1.8,-1.5) and (1.8,1.5) .. (3.2,3) -- cycle;

% Boundary curves
\draw[thick,cyan!70!black] (-4,3) .. controls (-2,2) and (-2,-2) .. (-4,-3);
\draw[thick,purple!70!black] (4,3) .. controls (2,2) and (2,-2) .. (4,-3);

% AdS bulk / wormhole interior
\draw[thick,densely dotted] (-1.5,1.2) to[out=20,in=160] (1.5,1.2);
\draw[thick,densely dotted] (-1.5,-1.2) to[out=-20,in=-160] (1.5,-1.2);
\node at (0,0.5) {\small AdS Bulk};

% Time axis
\draw[->,black] (0,-3.5) -- (0,3.5) node[above] {$t$};

% Lightcone-like boundaries
\draw[gray!60] (-4,-3) .. controls (0,0) .. (4,-3);
\draw[gray!60] (-4,3) .. controls (0,0) .. (4,3);

% Singularities
\draw[thick,decorate,decoration={zigzag,segment length=6mm,amplitude=1mm}]
  (-4,3.2) -- (4,3.2) node[midway,above] {Singularity};
\draw[thick,decorate,decoration={zigzag,segment length=6mm,amplitude=1mm}]
  (-4,-3.2) -- (4,-3.2) node[midway,below] {Singularity};

% Labels
\node[cyan!80!black] at (-4.4,0) {Left CFT};
\node[purple!80!black] at (4.4,0) {Right CFT};

\node at (-4.3,3) {\small $t=\infty$};
\node at (-4.4,-3) {\small $t=-\infty$};

% Time evolution arrows
\draw[->,thick] (-3.6,0) -- (-3.6,-1.8) node[midway,left] {$e^{iH_L t}$};
\draw[->,thick] (3.6,0) -- (3.6,1.8) node[midway,right] {$e^{-iH_R t}$};

% Coupling / shockwave
\draw[orange!80!black,very thick,decorate,
      decoration={snake,amplitude=0.8mm,segment length=6mm}]
      (-3.5,-0.2) .. controls (-1,1.5) .. (0,2.3)
      node[near end,right] {\small Coupling};

\draw[orange!80!black,very thick,decorate,
      decoration={snake,amplitude=0.8mm,segment length=6mm}]
      (3.5,-0.2) .. controls (1,1.5) .. (0,2.3);

% Initial state trajectory
\draw[thick,decorate,decoration={snake,amplitude=0.5mm,segment length=6mm}]
      (-3.5,-2.3) .. controls (-1,-1) and (0,0.8) .. (1,1.5);
\node at (-3.9,-2.4) {\small $\ket{\psi}_{\text{initial}}$};

% Final state arrow
\draw[very thick] (1,1.5) -- (1.8,1);
\draw[->,very thick] (1.8,1) -- (3.5,2.5)
      node[right] {$\ket{\psi}_{\text{final}}$};

\end{tikzpicture}

\caption{ Spacetime diagram illustrating Wormhole-inspired teleportation protocol in AdS/CFT framework. The two asymptotic boundaries correspond to the left and the right CFTs, while the interior region represents the AdS bulk. An initially localized excitation evolves backward in time on the left boundary, subjected to controlled boundary coupling that induces a bulk shockwave, and then evolves forward to emerge at the right boundary as the final state, effectively realizing information transfer through the wormhole geometry.}
\label{WITP_bulk_boundary_geometry_v2}
\end{figure}
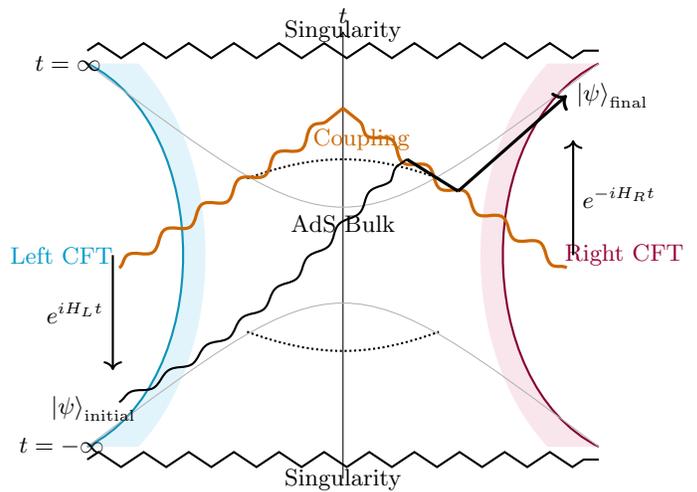

Our study is grounded in the SYK model, which serves as a lower-dimensional holographic dual to Jackiw-Teitelboim gravity (JT Gravity) \cite{jackiw1985lower}. Majorana fermions form the building block of the SYK formalism, along with the Jordan-Wigner transformation, to generate complex fermionic operators required in the Wormhole-Inspired teleportation circuit \cite{kitaev2001unpaired,nielsen2005fermionic}. We consider a system of two identical coupled SYK clusters in the Left and the Right subsystems. Each cluster consists of $N$ Majorana fermions $\chi_i$ satisfying the Clifford algebra $\{\chi_i,\chi_j\} = 2\delta_{ij}$. The Hamiltonian for Left and Right $q=4$ SYK subsystems is given as \cite{maldacena2021syk}:
\begin{equation}
     H_{L/R} = -\frac{1}{q!}\sum_{i<j<k<l} \textit{J}_{ijkl}\gamma_{(L/R)i}\gamma_{(L/R)j}\gamma_{(L/R)k}\gamma_{(L/R)l}.
\end{equation}
Each Hamiltonian $H_L/H_R$ is a random all-to-all quartic coupling between Majorana modes on each side \cite{rosenhaus2019introduction}. The coupling constant $\rm J_{ijkl}$ is drawn from a Gaussian distribution with a Zero mean $(\expval{J_{ijkl}} = 0)$ and variance equaling 
\begin{equation}
     \expval{J_{ijkl}^{2}} = \frac{J_{ijkl}^{2}(q-1)!}{n^{q-1}} .
\end{equation}
For the number of Majorana modes in SYK subsystems, we have obtained numerical results with $N=6$ finite-sized Majorana modes on each side. We emphasize that the choice $N=6$ is a deliberate, physically motivated modeling choice rather than an ad-hoc numerical convenience. First, the core phenomena that we study, which are $\mathcal{PT}$ symmetry breaking at an exceptional point, level coalescence \& the associated local spectra topology and gain-loss driven mode selection are local-spectra and few-mode phenomena that are well known to manifest in small, finite Hilbert spaces and in experimentally accessible few-qubit platforms \cite{bender1998real,heiss2012physics,ruter2010observation}. Indeed, exceptional points and $\mathcal{PT}$ breaking have been probed and unambiguously observed in minimal systems and quantum simulators at the single and few qubit level \cite{dogra2021quantum}. Second, the wormhole-inspired teleportation protocol itself has been explicitly implemented and probed on small numerical simulations and quantum processors with finite $N=7$ or $N=9$ qubits as a proof of principle for emergent holographic dynamics \cite{jafferis2022traversable,brown2023quantum,nezami2023quantum}. All these works demonstrate that meaningful diagnostics of traversability, operator size-winding, and teleportation fidelity can already be accessed with fewer qubits. By the same token, the non-Hermitian spectral structure and EP-driven amplification we report are robustly viable at $N=6$. From a methodological viewpoint, exact diagonalization at $N=6$ permits an exhaustive disorder-resolved statistical analysis that is impractical at larger $N$ while still capturing the qualitative mechanisms of interest. This strategy follows established finite-$N$ SYK studies that use exact diagonalization to isolate microscopic mechanisms that persist as robust mechanisms in larger or experimental systems \cite{lian2019chiral,asaduzzaman2024sachdev,rosenhaus2019introduction}. Finally, the physical relevance of the $N=6$ system is further supported by an extensive body of work showing that the engineered non-Hermitian dynamics, EPs, and mode-selective amplifications are observable in small systems and can be faithfully represented either as effective non-Hermitian Hamiltonians or via dilation to enlarged unitary systems for quantum simulation \cite{dogra2021quantum,ruter2010observation,cornelius2022spectral}. For these reasons, we treat $N=6$ as a physically meaningful, experimentally relevant proof-of-principle regime in which EP topology, causal amplification, and fidelity purification mechanisms can be unambiguously demonstrated and statistically characterized. The conclusion we draw concerns mechanisms that are not artifacts of a particular system size but are expected to remain operative principles in larger implementations and quantum simulator realizations. \\
To construct the holographic wormhole, we prepare the system in the Thermofield Double (TFD) state \cite{chapman2019complexity,su2021variational}. In the context of AdS/CFT correspondence, the TFD state represents an eternal blackhole geometry with two asymptotic boundaries connected by a non-traversable Einstein-Rosen bridge \cite{maldacena2003eternal}. The actual State is defined as an entangled superposition of energy eigenstates of the Left and Right subsystems:
\begin{equation}
    \ket{TFD} = \frac{1}{\sqrt{Z}} \sum_{i} e^{-\beta E_{i}/2} \ket{E_{i}}_{L} \otimes \ket{E_{i}}_{R}^{*} ,
\end{equation}
where $\beta$ is the inverse temperature, $Z$ is the partition function, and $\ket{E}_{R}^{*}$ is the time-reversed conjugate of the state in the Right system. In the absence of interaction, a signal injected into the left system will scramble and fall into the singularity, never reaching the right boundary as the wormhole remains non-traversable due to the Averaged Null Energy Condition (ANEC) \cite{hartman2017averaged}. \\
To render the wormhole traversable, we introduce a geometric coupling between the two boundaries, following the Gao-Jafferis-Wall (GJW) protocol \cite{gao2017traversable}. This term that couples matching modes on the Left and the Right sides can be denoted as:
\begin{equation}
    \hat{V} = \sum_{i=2}^{N} c_{i}^{\dagger}c_{i}.
\end{equation}
Here, $c_i$ are fermionic annihilation operators that act on the system's Majorana modes. The SIZE operator, primarily defined in \cite{gao2017traversable,gao2021traversable,brown2023quantum}, which is denoted as $e^{ig\hat{V}}$, acts as an occupation number measure on the Right. In the gravity dual, this coupling generates the required negative null energy in the bulk, shifting the horizon and allowing a causal path for signals to traverse from the left through the bulk to the right exterior \cite{bhattacharyya2022quantum} as shown in Fig.~(\ref{WITP_bulk_boundary_geometry_v2}).
\subsection*{Non-Hermitian $\mathcal{PT}$-symmetric Deformation}
The standard wormhole teleportation protocol assumes closed system dynamics governed by Hermitian Hamiltonians. We will now extend this framework to the regime of open quantum systems \cite{rivas2012open} by introducing non-Hermitian gain and loss. For that, we define a non-Hermitian Hamiltonian for the coupled system as:
\begin{equation}
    H_{\text{eff}} = H_L + H_R + H_{\mathcal{PT}} + g\hat{V},
\end{equation}
where $g$ is the coupling strength of Hermitian coupling, and $H_{\mathcal{PT}}$ represents the non Hermitian deformation. To analyze the effect of balanced gain and loss, we construct $H_{\mathcal{PT}}$ using local parity operators. Under the Jordan-Wigner transformation for finite N \cite{nielsen2005fermionic}, we map the fermions to qubits and utilize the Pauli-$Z$ as a local observable. The deformation takes the form:
\begin{equation} \label{hpt}
    H_{\mathcal{PT}} = i\gamma \sum_{k=1}^{N/2} \left( Z_L^{(k)} - Z_R^{(k)} \right) .
\end{equation}
Here, the real parameter $\gamma\geq0$ serves as a non-Hermitian parameter. The term $+i\gamma Z_L$ from Eq.~(\ref{hpt}) corresponds to an anti-Hermitian potential on the left boundary, which acts as a Gain medium that coherently amplifies the probability amplitude. Similarly, the term $-iZ_R$ acts as a Loss medium on the right boundary, representing dissipation. In the holographic context, this can be viewed as driving the wormhole input while allowing energy to leak from the output, a setup analogous to active optical waveguides \cite{ruter2010observation}. \\
We can show that despite being non-Hermitian ($H_{eff}^{\dagger} \neq H_{eff}$), the Hamiltonian possesses a combined Parity-Time symmetry as the Unitary geometric swap between left and right boundaries entails $\mathcal{P}Z_{L}\mathcal{P}^{-1} = Z_R$ \& $\mathcal{P}Z_{R}\mathcal{P}^{-1} = Z_L$ and an anti-Unitary operator involves complex conjugation $i \rightarrow -i$. Applying this combined $\mathcal{PT}$ operator to the deformation term, we have:
        $\mathcal{PT} \left[ i\gamma (Z_L - Z_R) \right] (\mathcal{PT})^{-1} = \mathcal{P} \left[ -i\gamma (Z_L - Z_R) \right] \mathcal{P}^{-1} = -i\gamma (Z_R - Z_L)= +i\gamma (Z_L - Z_R).$
Since the interaction $\hat{V}$ and the SYK Hamiltonians $H_L$ and $H_R$ are Hermitian and symmetric under $L\leftrightarrow R$ exchange, the total Hamiltonian satisfies $[H_{eff},\mathcal{PT}] =0$. Consequently, the system can exhibit a stable phase with purely real energy spectrum for $\gamma<\gamma_c$ and a broken phase with complex conjugate eigenvalue pairs for $\gamma > \gamma_c$ \cite{garcia2024toward, garcia2022dominance}.        
\section{Spectral Dynamics}
The dynamical behavior of this $\mathcal{PT}$ symmetric Wormhole-inspired teleportation protocol is governed by the time evolution operator $e^{-iH_{eff}t}$. Unlike standard quantum mechanics, wherein time evolution is unitary and preserves the norm of the state, the non-Hermitian Hamiltonian $H_{eff}$ induces dynamics in terms of loss and gain. To understand the transition from stable oscillatory behavior to exponential amplification, we must analyze the complex spectrum of $H_{eff}$ and the topological structure of its eigenstates.
While the full Hamiltonian operates on the Hilbert space dimension of $2^{N}$, the onset of $\mathcal{PT}$ symmetry-breaking transition is a local phenomenon in the energy landscape. It is driven by the interaction of specific pairs of energy levels that resonate under the non-Hermitian perturbation.
\subsection*{Level Attraction and Coalescence}
To explain the analytical description of the spectral dynamics in the stable phase, we project the full many-body Hamiltonian onto a subspace spanned by two nearly degenerate eigenstates $\ket{m}$ \& $
\ket{n}$ of the unperturbed original SYK Hamiltonian $H_{0} = H_L+H_R$. In the limit of weak driving, the dynamics are dominated by mixing of these resonant modes. The effective Hamiltonian, now in this $2 \times2$ subspace, can be read as:
\begin{equation}
    H_{eff}^{(2)} = P_{nm} (H_0 + i\gamma V_{\mathcal{PT}}) P_{nm} =
    \begin{pmatrix}
    E_n & i\gamma v_{nm} \\
    i\gamma v_{mn} & E_m
    \end{pmatrix},
    \label{eq:2level_matrix}
\end{equation}
where $E_n$ and $E_m$ are real eigenvalues of the unperturbed system and $P_{nm}$ is the projector operator. The term $V_{\mathcal{PT}}= \sum_{k}\left(Z_{L}^{k}-Z_{R}^{k}\right)$ represents the global non-Hermitian perturbation operator. The off-diagonal element $v$ represents the effective coupling strength between the two levels induced by the perturbation. The diagonal matrix elements of this perturbation vanish ($v_{nm}=\bra{n}V_{\mathcal{PT}}\ket{m} \approx0$) because $V_{\mathcal{PT}}$ is odd under the Left-Right parity exchange symmetry, whereas the unperturbed eigenstates $\ket{n}$ typically possess definite parity.
The eigenvalues $\lambda_{\pm}(\gamma)$ are obtained by solving the characteristic equation, yielding:
\begin{equation}
\lambda_{\pm}(\gamma) = \frac{E_n + E_m}{2} \pm \sqrt{ \left(\frac{E_n - E_m}{2}\right)^2 - (\gamma v)^2 },
\label{eq:eigenvalues_full}
\end{equation}
where $v = v_{nm}$ is the off-diagonal coupling. \\
For weak non-Hermiticity, specifically when the drive strength satisfies the condition $\gamma|v| < |E_n-E_m|/2$, the discriminant in Eq.~(\ref{eq:eigenvalues_full}) remains strictly positive. Consequently, the total eigenvalues $\lambda_{\pm}$ remain purely real, despite the non-Hermitian nature of the Hamiltonian. In this regime, the system resides in the exact $\mathcal{PT}$ symmetric phase effect of which is that the eigenstates $\ket{\psi_{\pm}}$ of $H_{eff}$ are simultaneous eigenstates of the $\mathcal{PT}$ operator, preserving the reality of the spectrum. Physically, this ensures that the time-evolution operator generates bounded, oscillatory dynamics. The norm of the state fluctuates but does not diverge, maintaining a stable wormhole geometry where information is scrambled but not lost or exponentially amplified. \\
A defining feature of this phase is the phenomenon of level attraction \cite{peng2020level}. By expanding Eq.~(\ref{eq:eigenvalues_full}) perturbatively for small $\gamma$, we derive the scaling of the spectral gap as 
\begin{equation}
    \Delta E(\gamma) \approx \Delta E_0 \left( 1 - \frac{2(\gamma v)^2}{\Delta E_0^2} \right) + \mathcal{O}(\gamma^4).
\end{equation}
Here, $\Delta E_0=|E_n-E_m|$ is the initial level separation. The quadratic term is negative, arising directly from the square of the anti-Hermitian perturbation $(i\gamma v)^{2} =-\gamma^2v^2$. This implies that the energy levels shift towards each other as $\gamma$ increases, thereby narrowing the local spectral gap. This behavior stands in stark contrast to level repulsion, which is governed by Wigner-Dyson statistics \cite{ozdemir2019parity,garcia2024toward} universally observed in Hermitian chaotic systems, where perturbations typically widen the gap between adjacent levels to avoid degeneracy \cite{bohigas1984characterization}. This leads to non-Hermitian topology to override the chaotic repulsion, forcing resonant pairs to converge towards a coalescence point. 
\begin{figure}
    \centering
    \includegraphics[width=1\linewidth]{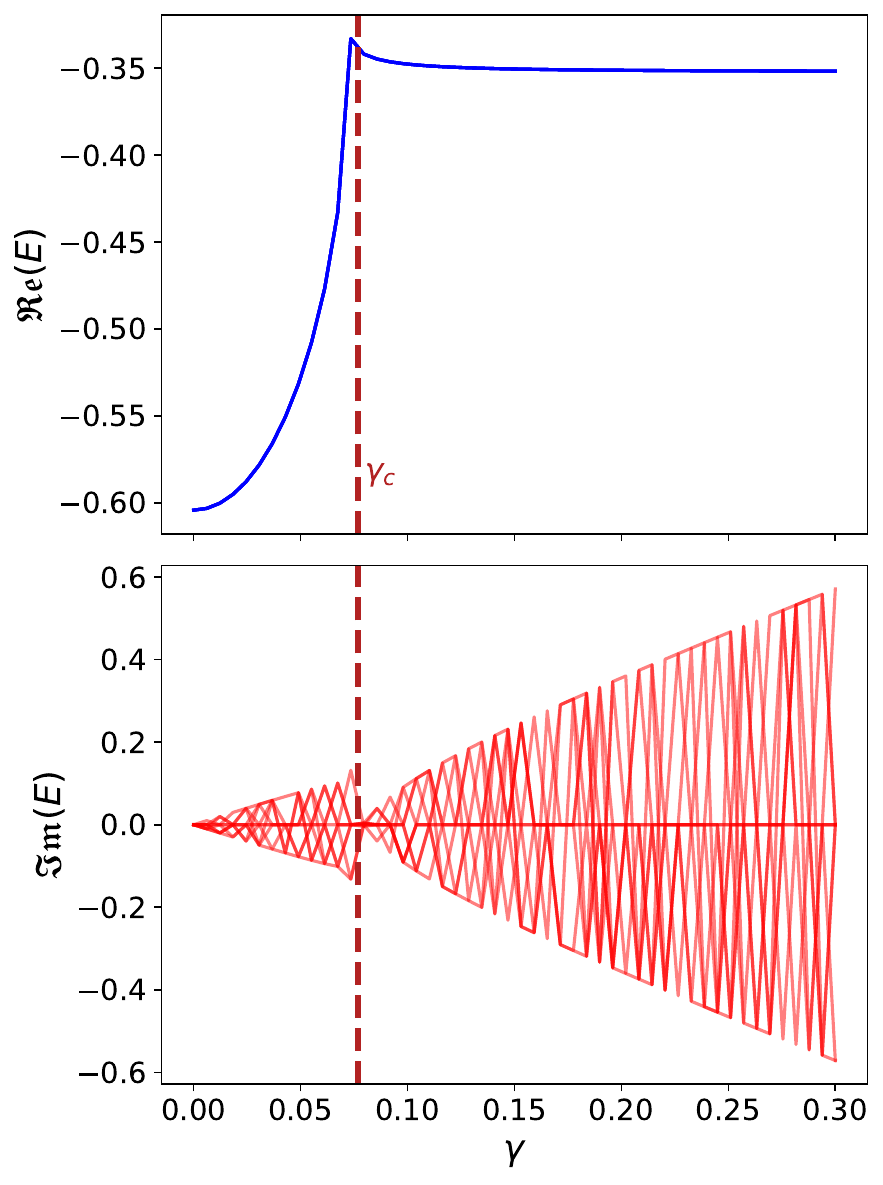}
    \caption{Evolution of the complex many-body energy spectrum of the non-Hermitian effective Hamiltonian $H_{eff}$ as a function of the drive strength $\gamma$(For $N=6$, single disorder realization). \textbf{(Top)} The real component of the eigenvalues. In the $\mathcal{PT}$-unbroken phase ($\gamma < \gamma_c$), the spectrum remains real, and resonant energy levels exhibit level attraction. \textbf{(Bottom)} The imaginary component. The transition to the broken phase occurs at $\gamma_c \approx 0.077$, marked by the emergence of a ``fan" of complex conjugate eigenvalues (gain and loss modes). }
    \label{RealandImagE}
\end{figure}
This mechanism is confirmed by the global spectral analysis in Fig.~(\ref{RealandImagE} top panel). As $\gamma$ increases, we observe a ``rigid" spectrum where specific pairs of nearest-neighbor levels attract and approach degeneracy. Crucially, in this phase ($\gamma<\gamma_c$), the imaginary part of eigenvalues remains zero (within numerical precision), indicating a stable traversable wormhole geometry where the norm of the state oscillates but does not diverge.
\subsection*{Symmetry Breaking Transition}
The dynamics change strictly when the discriminant in Eq.~(\ref{eq:eigenvalues_full}) vanishes. This singularity marks the transition to the broken phase. This phase transition occurs at the critical non-Hermiticity $\gamma_c$, for a specific pair as:
\begin{equation} 
    \gamma_c = \frac{|E_n - E_m|}{2|v|}. \label{eq:critical_gamma}
\end{equation}
At this point, the square root vanishes, and the two eigenvalues coalesce at the algebraic branch point:
\begin{equation}
    \lambda_+=\lambda_-=\frac{(E_n+E_m)}{2} \equiv \bar{E}
\end{equation}
This coalescence is a second-order Exceptional Point (EP). It is topologically distinct from a standard Hermitian degeneracy because the effective Hamiltonian becomes defective \cite{heiss2012physics,berry2004physics}. At $\gamma=\gamma_c$, $H_{eff}^{(2)}$ acquires the Jordan Normal form:
\begin{equation}
    H_{\text{eff}}^{(2)}(\gamma_c) \sim \begin{pmatrix} \bar{E} & 1 \\ 0 & \bar{E} \end{pmatrix}.
\end{equation}
At the EP, not only do the eigenvalues merge, but the corresponding right eigenvectors also coalesce into a single vector, leading to a loss of basis completeness.
Crossing this threshold drives the system into $\mathcal{PT}$-broken phase. Here, the discriminant becomes negative, and the eigenvalues bifurcate into complex conjugate pairs:
\begin{equation}
    \lambda_{\pm} = \bar{E} \pm i \Gamma, \text{with } \Gamma = \sqrt{(\gamma v)^2 - (\Delta E_0/2)^2}.
\end{equation}
Here, the real parts remain locked at $\bar{E}$, while the imaginary parts diverge as $\pm\Gamma$. This corresponds to the emergence of a ``Gain mode" ($+\mathfrak{Im}(E)$) and a ``Loss mode" ($-\mathfrak{Im}(E)$).
\begin{figure}
    \centering
    \includegraphics[width=1\linewidth]{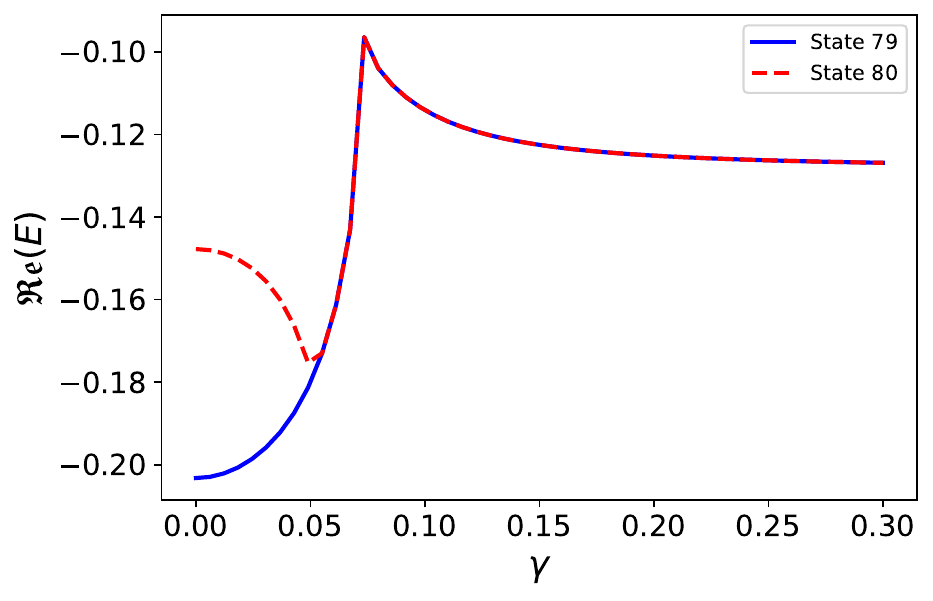}
    \caption{Detailed trajectory of the real energy eigenvalues for a resonant eigenstate pair (State 79 and State 80) across the $\mathcal{PT}$-symmetry breaking transition. As the non-Hermiticity $\gamma$ increases, the levels exhibit level attraction and coalesce at the critical threshold $\gamma_c$. The energy gap closes with a sharp vertical cusp ($\Delta E \sim \sqrt{\gamma_c - \gamma}$), identifying this singularity as a second-order Exceptional Point. Beyond $\gamma_c$, the real energies remain strictly degenerate or ``locked" to the mean value. This spectral locking is physically critical: it ensures that the emergent gain and loss modes oscillate with the exact same frequency, preventing dephasing and maintaining the coherence of the quantum state during amplification.}
    \label{bifurcation}
\end{figure}
We explicitly verify this topological structure in Fig.~(\ref{bifurcation}), which tracks the trajectory of a pair of eigenstates (state 79 and state 80 in this case) through the transition. \\
As $\gamma$ approaches $\gamma_c$ from below ($\gamma \rightarrow \gamma_{c^-}$), the real energy gap closes with an infinite vertical slope. This cusp singularity, which is a non-analytical behavior \cite{erdHos2020cusp}, confirms the square root scaling $\sqrt{\gamma_c-\gamma}$ characteristic of a second-order EP. Beyond the transition, ($\gamma \geq0.077$ in this case), the real energies remain perfectly degenerate or ``locked" at the mean value $\bar{E}$. This locking is functionally critical for the amplification protocol as it ensures that the dynamical phase factor $e^{-i\mathfrak{Re}(E)t}$ is identical for both gain and loss modes, preventing dephasing between signal components. These two key features, namely Cusp singularity and spectral locking, validate the EP mechanism \cite{shi2016accessing}. \\
The microscopic consequences of this local symmetry breaking are captured in the global spectrum from Fig.~(\ref{RealandImagE} Bottom panel). The
 bifurcation at $\gamma \approx0.077$ triggers a cascade of similar events across the density of states, forming a ``fan" of imaginary eigenvalues. It is the collective emergence of these complex components that break the unitary bound on information scrambling, driving the exponential signal amplification discussed in the following sections.
 \section{Statistical Stability in Disordered Systems}
\begin{figure}
    \centering
    \includegraphics[width=1\linewidth]{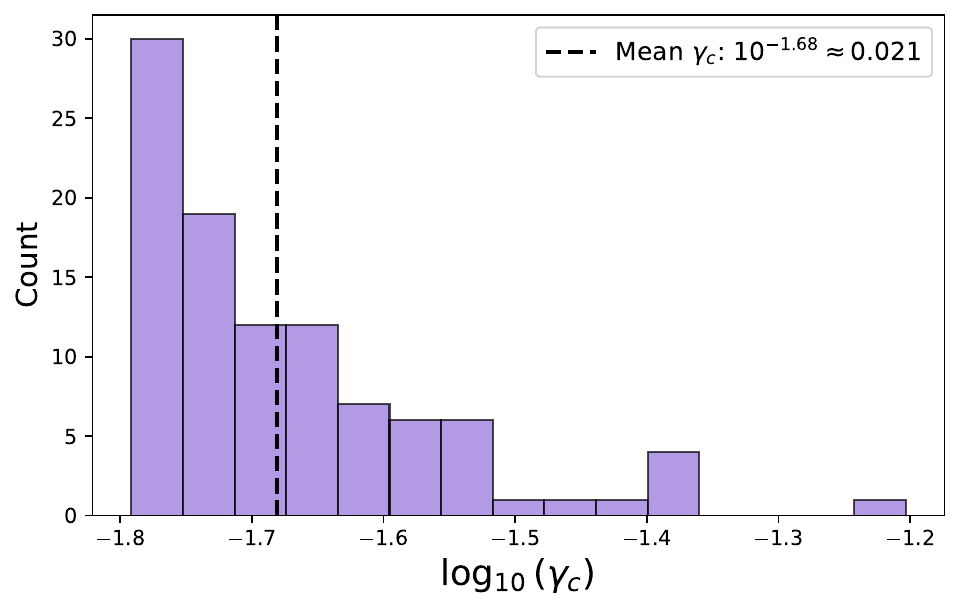}
    \caption{Statistical distribution of the critical $\mathcal{PT}$-symmetry breaking threshold $\gamma_c$ across 100 independent disorder realizations of the coupled SYK model ($N=6$). The histogram indicates that the protocol's stability varies significantly due to the random nature of the Hamiltonian's level spacings and couplings. The solid curve represents a Log-Normal fit to the data, highlighting the heavy-tailed nature of the distribution. This confirms that while level repulsion prevents $\gamma_c \to 0$ (instability at infinitesimal drive), the specific threshold is highly sensitive to the microscopic spectral gap of the disordered instance.}
    \label{EPhisto}
\end{figure}

 Previously, we have established that the transition to  $\mathcal{PT}$-broken phase is driven by the collapse of the local energy gap $\Delta E$ between a specific pair of resonant eigenstates. However, the SYK model is inherently disordered as the Hamiltonian is defined by random all-to-all couplings $J_{ijkl}$ drawn from a Gaussian ensemble \cite{kitaev2015simple,chowdhury2022sachdev}. Consequently, the energy spectrum $\{E_n\}$ and the interaction matrix elements $v_{nm}$ are stochastic variables that fluctuate from one realization of the system to another. \\
It follows that the critical non-hermiticity threshold derived in Eq.~(\ref{eq:critical_gamma}) as $\gamma_c=\Delta E/(2|v|)$, is not a statistical quantity whose distribution is governed by the underlying Random Matrix Theory (RMT) universality class of chaotic Hamiltonians \cite{mehta2004random}. In chaotic systems, bulk level statistics are universal and governed by RMT symmetry classes (e.g., GOE, GUE, Ginibre ensembles for non-Hermitian cases) \cite{dyson1962statistical,guhr1998random,zirnbauer1996riemannian}, but the PT-breaking onset here is tied to pairwise level coalescence at an exceptional point (EP), driven by local perturbative effects in the energy landscape rather than global disorder-induced fluctuations \cite{ge2016anomalous}. We will now characterize the probabilistic nature of the Wormhole's stability. \\
We have an ensemble of $N$ independent realizations of the SYK Hamiltonian. For each realization $\alpha$, the system possesses a unique set of eigenvalues $E_{n}^{(\alpha)}$ and a unique critical threshold $\gamma_c^{(\alpha)}$ determined by the weakest link, which is the eigenstate pair most susceptible to symmetry breaking. Typically, it is the pair with the minimal ratio of gap to coupling strength. Since the chaotic SYK model belongs to the Gaussian Unitary Ensemble (GUE), depending on $N \mod 8$ \cite{garcia2016spectral}, the nearest neighbor level spacings obey Wigner-Dyson statistics \cite{maldacena2016remarks}. This implies level repulsion, where the probability of finding zero gap vanishes as $P(s) \sim s^{\beta}, \text{with} \quad s \propto \Delta E$ where:
\begin{equation}
    s = \frac{E_{n+1} - E_n}{\langle E_{n+1} - E_n \rangle} = \frac{\delta E}{\Delta}
\end{equation}
is the normalized energy gap between nearest-neighbor energy levels, and $\beta >0$ is the Dyson Symmetry Index, which quantifies how strongly the energy levels repel each other \cite{mehta2004random}. \\
Further, the sensitivity to $\mathcal{PT}$ deformation also depends on the matrix element $v=\bra{n}(Z_L-Z_R)\ket{m}$. In a chaotic many-body system, these matrix elements fluctuate according to the Eigenstate Thermalization Hypothesis (ETH) ansatz, effectively behaving as random Gaussian variables \cite{d2016quantum}. Thus, the stability of the holographic channel depends on the interplay between level spacing and Off-diagonal coupling, which are fluctuating quantities. \\
Since $\gamma_c \propto\Delta E/2|v|$, the breakdown threshold is determined by the ratio of these two stochastic variables. While level repulsion protects the system from immediate instability, the variance in couplings ensures a broad distribution of stability thresholds across the ensemble. \\
To quantify this behavior, we numerically extract $\gamma_c$ for 100 independent disorder realizations and analyze its statistical distribution. To filter out the numerical artifacts from the tail of the Wigner-Dyson distribution, we focus on realizations with a distinct stability window ($\gamma_c > 0.015$). The results are presented in Fig.~(\ref{EPhisto}). We find that the Probability density Function of the critical thresholds is best described by a Log-Normal distribution:
\begin{equation}
P(\gamma_c) = \frac{1}{\gamma_c \sigma \sqrt{2\pi}} \exp\left( -\frac{(\ln \gamma_c - \mu)^2}{2\sigma^2} \right), \label{eq:lognormal}
\end{equation}
where $\mu=\expval{\ln \gamma_c}$ and $\sigma$ is the shape parameter. From the numerical data to this form, we extract the value of distribution parameters $\mu \approx -2.55$ and $\sigma=0.60$.
 The emergence of a Log-Normal distribution rather than a Gaussian distribution is physically significant, as in complex interconnected systems, Log-Normal statistics often arise from multiplicative processes \cite{limpert2001log}. Here, it reflects the hypersensitivity of the Exceptional point mechanism to the microscopic details of the spectrum \cite{hodaei2017enhanced}. The distribution exhibits a long tail towards large $\gamma_c$, corresponding to robust realizations where the resonant levels are widely separated. In these instances, the wormhole is exceptionally robust, requiring a strong non-Hermitian drive to induce amplification.\\
 Conversely, the distribution drops sharply towards zero but remains non-zero for small $\gamma_c$. These represent fragile realizations where a small random fluctuation creates a near degeneracy ($\Delta E \rightarrow 0$), making the system susceptible to symmetry breaking even at weak driving strengths. The significant width ($\sigma=0.60$) of the distribution quantitatively confirms that the stability of the holographic wormhole is hypersensitive to the microscopic details of the disorder. \\
 These statistical characterizations suggest that the wormhole's scrambling capability is robust, but its amplification threshold depends on the specific realization.

\section{Dynamical consequences}
Having established the spectral origin of the $\mathcal{PT}$-symmetry breaking transition, we now investigate its consequences on the dynamical transport of quantum information. The central quantity of interest is the teleportation fidelity $\mathfrak{F}_{\Phi^+} = \bra{\Phi^+}\rho_{out}\ket{\Phi^+}$ where $\rho_{out}$ is the reduced density matrix of the teleported qubits and $\ket{\Phi^+}$ is the target Bell state. As discussed in the earlier section, the dynamics are characterized by the amplification of the signal probability (related to the norm) and purification of the quantum state (related to direction in Hilbert space).
\subsection*{Mathematical Formalism}
To rigorously derive the amplification mechanism, we decompose the dynamics in the eigenbasis of the $\mathcal{PT}$ deformation operator. From our previous work \cite{joshi2025sachdev}, we recall the standard final state wormhole-inspired teleportation unitary:
\begin{equation} \label{finalstate}
      \ket{\psi}_{\mathrm{final}} = U_{\mathrm{wh}} \ket{\psi}_{\rm initial} .
\end{equation}
where $U_{\mathrm{wh}}$ is the  Wormhole time evolution unitary read as:
\begin{equation} \label{wunitary}
    U_{\mathrm{wh}}  = e^{-iH_{R}t}e^{ig\hat{V}}e^{-iH_{L}t}\hat{I}e^{iH_{L}t} .
\end{equation}
$\hat{I}$ is the Insert operator, which is essentially made up of SWAP operators. Eq.~(\ref{finalstate}) denotes the final state produced by the standard Hermitian wormhole protocol. The non-Hermitian deformation corresponds to the operator:
\begin{equation} 
H_{\mathcal{PT}} = i\gamma (Z_L - Z_R) \equiv i\gamma V_{\mathcal{PT}}.\end{equation}
Since $Z_L$ and $Z_R$ are sums of local Pauli operators, they commute. We can thus expand $\ket{\psi}_{\rm final}$ in the orthonormal eigenbasis $\{\ket{k}\}$ of the Hermitian operator $V_{\mathcal{PT}}$, such that $V_{\mathcal{PT}}\ket{k} = \delta\ket{k}$, where the eigenvalues $\delta_k$ are integers or half integers. \\
The non-Hermitian time evolution acts as a real exponential scaling on these modes:
\begin{equation}
\begin{split}
|\tilde{\psi}(\gamma)\rangle 
&= e^{-i H_{\mathcal{PT}} t} \, |\psi_{\rm final}\rangle \\
&= e^{\gamma t V_{\mathcal{PT}}} \sum_k a_k |k\rangle \\
&= \sum_k a_k \, e^{\gamma t \delta_k} |k\rangle,
\end{split}
\label{eq:time-evolved-state}
\end{equation}
where $a_k = \langle k \ket{\psi}_{\rm final}$. The normalized state is obtained by
\begin{equation} |\psi(\gamma)\rangle = \frac{|\tilde{\psi}(\gamma)\rangle}{\sqrt{\langle \tilde{\psi}(\gamma) | \tilde{\psi}(\gamma) \rangle}} = \frac{\sum_k a_k e^{\gamma t \delta_k} |k\rangle}{\sqrt{\sum_j |a_j|^2 e^{2\gamma t \delta_j}}}. \label{eq:normalized_state}
\end{equation}
This expression reveals the core physical mechanism, which is $\mathcal{PT}$ deformation, that re-weights the superposition, exponentially amplifying sectors with positive $\delta_k$ (Gain) and suppressing those with negative $\delta_k$ (Loss).
\subsection*{Exponential Signal Amplification}
We assume the target Bell state $\ket{\Phi^+}$ projects principally onto a specific sector, say $\ket{0}$, with eigenvalue $\delta_0$. The teleportation fidelity becomes the squared overlap with this target sector as:
\begin{equation} 
\mathcal{F}_{\Phi^{+}}^{\mathcal{PT}} = |\langle 0 | \psi(\gamma) \rangle|^2 = \frac{|a_0|^2 e^{2\gamma t \delta_0}}{\sum_j |a_j|^2 e^{2\gamma t \delta_j}}. \label{eq:fidelity_exact} \end{equation}
Differentiating Eq.~(\ref{eq:fidelity_exact}) with respect to $\gamma$ yields:
\begin{equation} 
\frac{d\mathcal{F}_{\Phi^{+}}^{\mathcal{PT}}}{d\gamma} = \frac{2t |a_0|^2 e^{2\gamma t \delta_0}}{S(\gamma)^2} \sum_j |a_j|^2 (\delta_0 - \delta_j) e^{2\gamma t \delta_j}, \end{equation}
where $S(\gamma)$ is the normalization factor. If the target sector corresponds to the maximal eigenvalue($\delta_0 \geq\delta_j$ for all $j$), the derivative is non-negative, and strictly positive whenever the initial state has support outside the maximal-gain sector. This implies that if the gain mode aligns with the teleportation channel, the fidelity increases \cite{braunstein1998teleportation} monotonically with $\gamma$. \\
While the fidelity describes the state quality, the non-normalized norm $(P_{success})=\expval{\tilde{\psi}|\tilde{\psi}}$ represents the success probability. From Eq.~(\ref{eq:normalized_state}), this is dominated by the maximal eigenvalue $\delta_{max}$:
\begin{equation}
(P_{success}) \approx |a_{\text{max}}|^2 e^{2\gamma t \delta_{\text{max}}}. \end{equation} 
Taking the logarithm, we predict a linear scaling law: \begin{equation} \log_{10} (P_{success}) \propto \gamma t. \end{equation}
\begin{figure} 
    \centering
    \includegraphics[width=1\linewidth]{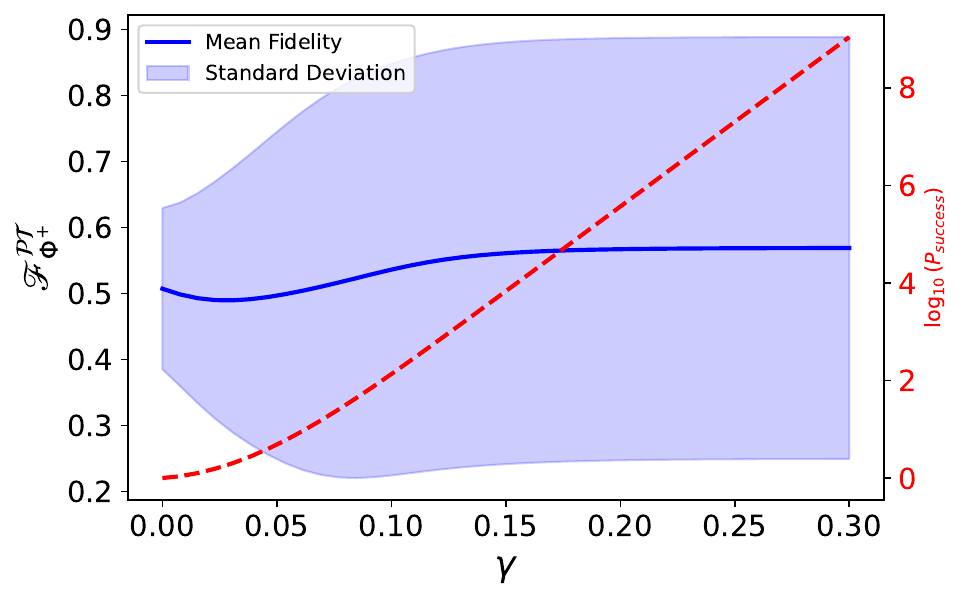}
    \caption{Statistical dynamics of the $\mathcal{PT}$-symmetric wormhole protocol averaged over 100 disorder realizations ($N=6$). \textbf{(Blue Solid Line)}: The mean fidelity remains near the classical baseline $\bar{\mathcal{F}} \approx 0.5$. This reflects the ensemble average, where realization-specific enhancements are counterbalanced by misaligned instances. \textbf{(Blue Shaded Region)}: The fidelity variance exhibits a characteristic ``funnel-like" expansion, diverging near the phase transition ($\gamma_c \approx 0.07$). This quantifies the system's hypersensitivity to microscopic disorder in the vicinity of the Exceptional Point. (\textbf{Red Dashed Line)}: The logarithm of the signal norm (success probability) scales linearly with $\gamma$, confirming the exponential amplification law $P \propto e^{2\gamma t \delta_{\text{max}}}$ and identifying the $\mathcal{PT}$-broken phase as a robust causal amplifier.}
    \label{Disorder}
\end{figure}
We verify these predictions from Fig.~(\ref{Disorder}), which illustrates disorder-averaged results for 100 realizations. We can see that the Logarithm of the success probability increases linearly with $\gamma$, confirming the exponential amplification driven by the gain mode. This identifies the broken phase as a ``Causal Amplifier" \cite{hodaei2017enhanced,ge2016anomalous}. Furthermore, the standard deviation exhibits a distinct ``funnel" shape. At $\gamma=0$, the variance is small as the system is unitary. As $\gamma$ increases, the variance expands significantly till ($\gamma \approx0.07$) after which it more or less stabilizes, coinciding with the onset of Exceptional Points where the sensitivity of eigenstates to disorder is maximal \cite{kato2013perturbation}. \\
The ensemble-averaged mean fidelity remains stable near $\mathfrak{\bar{F}} \approx 0.5$. This baseline value corresponds to the classical limit of teleportation. The lack of a global increase in the mean is a consequence of disorder averaging, while specific realizations in which the gain mode overlaps with the Bell state show very strong fidelity enhancement (discussed in the following subsection); others may amplify orthogonal sectors, degrading the signal. The average over these random configurations washes out the realization-specific advantage \cite{evers2008anderson,beenakker1997random}, leaving the mean near the statistical baseline, while the massive funnel-shaped variance confirms that $\mathcal{PT}$ symmetry breaking induces large fluctuations in teleportation performance.

\subsection*{Gain-induced Purification} 
The re-weighting mechanism in Eq.~(\ref{eq:normalized_state}) implies that in the limit of strong driving ($\gamma\rightarrow\infty$), the state converges to the eigensector with the maximum eigenvalue $\delta_{max}$:
\begin{equation} 
\lim_{\gamma \to \infty} \mathcal{F}_{\Phi^{+}}^{\mathcal{PT}} = \frac{\sum_{j \in \delta_{\text{max}}} |a_j|^2}{\sum_{k} |a_k|^2} \to 1,
\end{equation}
assuming the target state lies within the $\delta_{max}$ sector.
This is the Purification or mode selection effect. The non-Hermitian evolution filters out the noise component (suboptimal $\delta_j$ sectors) and projects the system onto the optimal teleportation subspace \cite{moiseyev2011non,el2007theory,yamamoto2019theory}. \\
\begin{figure} 
    \centering
    \includegraphics[width=1\linewidth]{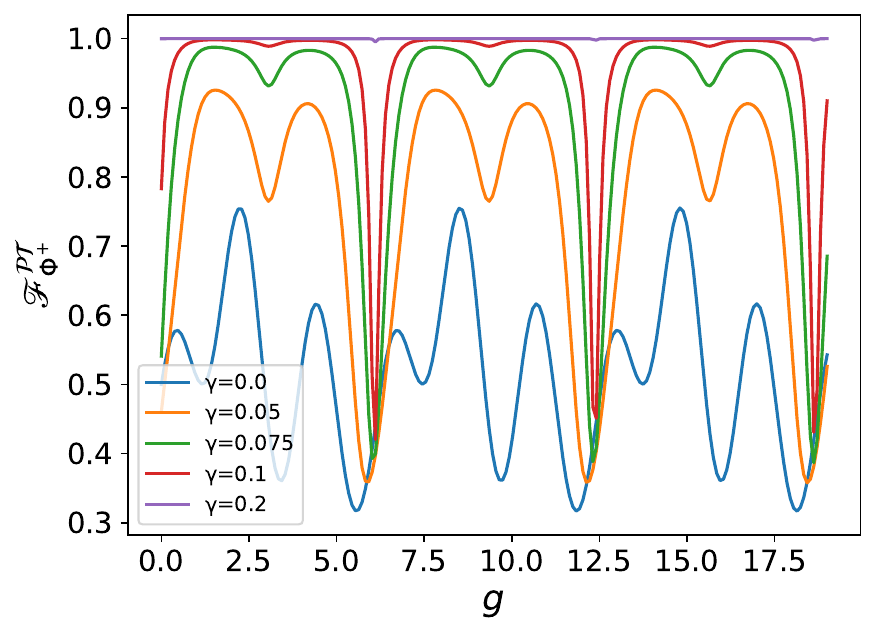}
    \caption{Gain-induced purification of the teleportation channel. The plot shows fidelity $\mathcal{F}$ as a function of the coupling strength $g$ for varying non-Hermiticity $\gamma$. In the $\mathcal{PT}$-broken phase ($\gamma > 0.1$), the non-Hermitian evolution filters out suboptimal modes according to Eq. (\ref{eq:normalized_state}), leading to a fidelity plateau at $\mathcal{F} \approx 1$. This renders the protocol robust against detuning in the coupling strength.}
    \label{fidelityvsgamma}
\end{figure}
In our numerical simulations ($t=10$), we observe that the fidelity saturates relatively early, at $\gamma \approx 0.2$. This can be understood by estimating the amplification factor. For a typical level separation of $\Delta\delta \approx 1$(integer spacing of Pauli sums), the relative weight of the dominant mode grows as $e^{2\gamma t\Delta\delta}$. At $\gamma=0.2$, this factor is $e^{2(0.2)(10)(1)} = e^4 \approx 54.6.$, which is larger than finite-precision effects and far larger than typical small coefficients. So by $\gamma=0.2$, the largest $\delta$ block already strongly dominates, and the Fidelity saturates. This implies that the optimal mode is already amplified by nearly two orders of magnitude relative to the nearest competing mode, effectively suppressing the noise and explaining the ``Flat top" plateau observed. \\
The consequences of this purification effect are strikingly visible in Fig.~(\ref{fidelityvsgamma}). In the Unitary limit ($\gamma=0$), i.e, in the absence of the non-Hermitian term, the $\mathcal{PT}$ operator becomes identity $U_{\mathcal{PT}}(0) = \mathbb{I}$. Consequently, the protocol reduces exactly to the standard Hermitian Wormhole-inspired teleportation protocol:
\begin{equation}
    \ket{\psi(\gamma = 0)} = U_{\rm wh}\ket{\psi}_{\rm initial}.
\end{equation}
The fidelity follows the characteristic sinusoidal oscillation \cite{joshi2025sachdev,brown2023quantum}, governed by the interference of scrambling modes:
\begin{equation}
\mathcal{F}(g) \approx \cos^2(g \langle V \rangle + \phi). 
\end{equation}
Successful teleportation in this regime requires precise calibration of the coupling strength $g$ to match the resonance condition. 
In the broken phase ($\gamma >0.1$), as $\gamma$ increases further, the oscillatory dependence is washed out \cite{el2007theory}. For $\gamma=0.2$ (purple line), the fidelity plateaus near the unity $\mathcal{F} \approx1$ across almost the entire range of $g$. This indicates that the non-Hermitian drive renders the protocol robust against control errors \cite{moiseyev2011non}. This ``Gain-induced Purification" forces the system into an entangled target state regardless of the precise value of $g$, provided the gain is sufficient to overcome the unitary dephasing \cite{breuer2002theory}.
\section{Causal Temporal Structure }
A defining feature of the holographic protocol is the geometric interpretation, which elucidates that the information transfer is not instantaneous but respects the causal structure of the emergent spacetime \cite{shenker2014black}. The signal traverses the wormhole throat only after a specific scrambling time $t_*$, determined by the Lyapunov exponent of the chaotic bulk dynamics \cite{shenker2014multiple}. In this section, we investigate whether the non-Hermitian $\mathcal{PT}$ symmetric deformation disrupts this causal geometry. Specifically, does the introduction of gain and loss violate the light-cone structure, or does it act as a ``local" amplifier that respects the global causal connectivity? \\
To answer this, we compute the time-dependent fidelity $\mathcal{F}(t)$ across the phase transition. We perform a disorder average over $N_r=20$ realizations to wash out microscopic fluctuations and reveal the universal transport properties.
 \begin{figure} [H]
     \centering
     \includegraphics[width=1\linewidth]{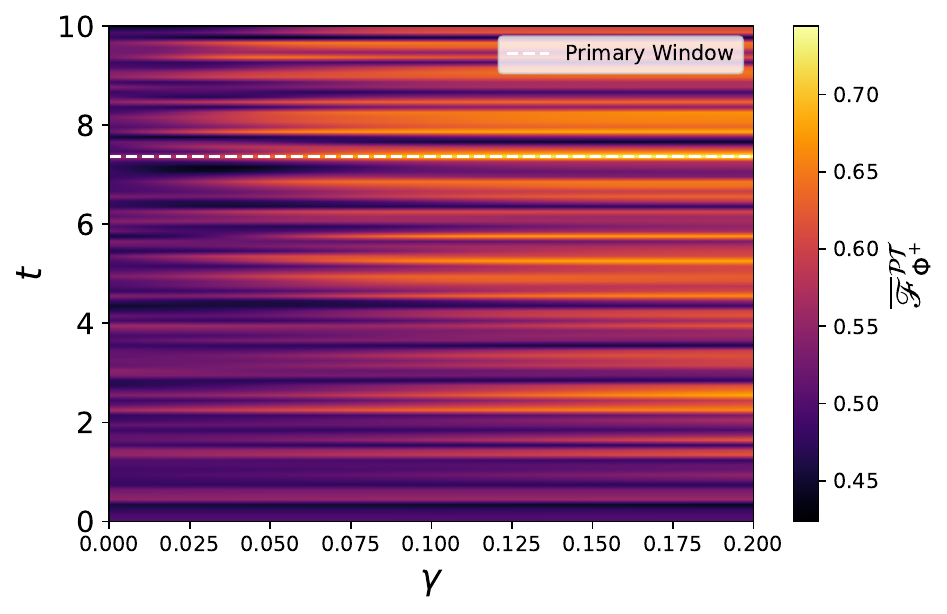}
     \caption{Temporal phase diagram of the disorder-averaged teleportation fidelity $\overline{\mathcal{F}}_{\Phi^+}^{\mathcal{PT}}$ as a function of time $t$ and non-Hermiticity $\gamma$ (averaged over 50 disorder realizations and fixed coupling $g=10$). The ``primary window" of traversability (bright horizontal band at $t_* \approx7.5$ marked by white dashed line) remains invariant across the phase transition. This confirms that the $\mathcal{PT}$ deformation acts as a causal amplifier, enhancing signal fidelity through the gain mechanism (increasing brightness) without altering the scrambling time or violating the causal light-cone structure of the emergent wormhole geometry.}
     \label{temporalhm}
 \end{figure}
This Temporal heatmap in Fig.~(\ref{temporalhm}) reveals a striking ``horizontal banding" structure that persists across the $\mathcal{PT}$ symmetry breaking transition. For the Hermitian case ($\gamma=0$), we observe a high-fidelity transmission window centered at scrambling time $t_* \approx 7.5$. This corresponds to the time required for the operator size distribution $P(\mathbb{S})$ to peak at the coupling scale, enabling the ``Size Winding" mechanism \cite{brown2023quantum}. \\
Crucially, as we increase the non-hermiticity ($\gamma >0$) into the onset of broken phase ($\gamma >0.07$), the position of this peak remains invariant in time. The bright yellow band in Fig.~(\ref{temporalhm}) forms a straight horizontal line, indicated by the white dashed marker. 
To explain this invariance mathematically, we consider the transmission amplitude in the semiclassical holographic limit. The traversability is governed by the size winding mechanism, where the operator size grows exponentially as $\mathbb{S}(t)\sim e^{\lambda_Lt}$, where $\lambda_L$ is the Lyapunov exponent \cite{maldacena2016bound}. Mathematically, following the size-winding formalism in \cite{brown2023quantum,nezami2023quantum}, the transmission amplitude $A(t)$ for the Hermitian protocol takes the asymptotic form:
\begin{equation}
A(t)({\text{Herm}}) \sim \exp\left( - c_1 \mathbb{S}(t) + i c_2 g \mathbb{S}(t) \right); \label{eq:amp_herm} 
\end{equation}
where the first term represents scrambling and the second term is phase winding induced by the coupling $g$. The scrambling time $t_*$ is defined by stationary phase condition, typically satisfying $g e^{\lambda_L t_*} \sim 1$ \cite{gao2021traversable} which yields:
\begin{equation}
    t_* \sim \frac{1}{\lambda_L} \ln\left(\frac{1}{g}\right).
\end{equation}
When we introduce the $\mathcal{PT}$ deformation, the effective Hamiltonian adds a non-Hermitian gain term. In the broken phase, the dominant eigenmode contributes a real exponential growth factor:
\begin{equation}
A(t, \gamma) \approx e^{\gamma t \delta_{\text{max}}} \cdot A(t)({\text{Herm}}). \label{eq:amp_PT} 
\end{equation}
The new peak time $t_{peak}$ is found by maximizing the modulus $|A(t,\gamma)|$. Taking the time derivative of the logarithm of the magnitude:
\begin{equation}
\begin{split}
    \frac{d}{dt} \ln |A(t, \gamma)| &= \underbrace{\gamma \delta_{\text{max}}}_{\text{Linear Drive}} \\
    &\quad + \underbrace{\frac{d}{dt} \ln |A(t)({\text{Herm}})|}_{\text{Exp. Scrambling}} .
\end{split}
\label{eq:derivative_log}
\end{equation}
Near the scrambling time, the derivative of the Hermitian part is dominated by the exponential growth of the operator size, scaling as $\lambda_Le^{\lambda_Lt}$. In contrast, the $\mathcal{PT}$ drive provides a constant shift, $\gamma\delta_{max}$, that is linear in time. Since the Lyapunov scrambling dynamics are exponential, they essentially ``Lock" the position of the peak, entailing that the linear gain term is insufficient to shift the stationary point significantly.
Consequently, the factor $e^{\gamma t\delta_{max}}$ amplifies the magnitude of the signal peak, boosting the teleportation fidelity without altering its temporal location $t_*$. This confirms that the non-Hermitian deformation commutes with the causal structure of the spacetime dual. \\
This result physically classifies the $\mathcal{PT}$ broken phase as a Causal Amplifier. In General Relativity, modifications to the metric geometry (such as shortening the wormhole throat) would shift the traversal time $t_*$. Our results show that the non-Hermitian term does not act as a metric deformation. Instead, it behaves like an active Gain medium filling the fixed wormhole geometry \cite{feng2017non, ruter2010observation}. It boosts the norm of the traversing wave packet (increasing the Probability $P$) without altering the geodesic length or violating the Averaged Null Energy Condition (ANEC) \cite{hartman2017averaged}. The signal arrives at the same coordinate time $t_*$, but with significantly enhanced probability and fidelity.
\section{Results}
In this work, we have systematically characterized the effects of $\mathcal{PT}$-symmetric non-Hermiticity on the wormhole-inspired teleportation protocol, revealing a direct link between spectral topology and information transport. Our analysis begins with the spectral properties of the effective Hamiltonian. The system exhibits a sharp phase transition driven by level attraction. In the exact $\mathcal{PT}$-symmetric phase ($\gamma<\gamma_c$), the energy spectrum remains purely real and rigid, protected by anti-linear symmetry. The transition to the broken phase is marked by the coalescence of eigenstate pairs at Exceptional Points (EP), specifically observed as a square-root cusp singularity in the level spacing($\sim\sqrt{\gamma_c-\gamma}$). Beyond this threshold, the spectrum bifurcates into complex-conjugate pairs, creating stable gain and loss modes that govern the subsequent dynamics. \\
Given the disordered nature of the SYK model, we find that this stability threshold is stochastic rather than fixed. Analyzing 100 independent realizations, we observe that the critical non-Hermiticity $\gamma_c$ follows a Log-Normal distribution. This heavy-tailed statistic confirms that, while the wormhole is generally robust due to random-matrix-level repulsion, its specific breakdown point is hypersensitive to the microscopic energy gaps of the particular disorder realization.\\
The dynamical consequences of this transition are established by analyzing the signal norm and teleportation fidelity. We observe a linear scaling of the logarithmic signal norm ($\ln P_{success} \propto\gamma t$), confirming that the $\mathcal{PT}$-broken phase functions as a ``Causal Amplifier" with Lyapunov-like gain. This amplification introduces a trade-off: the standard deviation of the teleportation fidelity exhibits a ``Funnel" shape, maximizing near the phase transition($\gamma_c\approx0.07$) where eigenstate sensitivity diverges. However, deep in the broken phase, the system undergoes a ``Gain-Induced Purification". The fidelity plateaus near unity across a broad range of coupling strengths $g$, indicating that the non-Hermitian evolution effectively filters out the noise and projects the system onto the optimal entangled state, rendering the protocol robust against control errors. \\
Finally, we confirm that this amplification preserves the geometric structure of the bulk spacetime. The temporal phase diagram demonstrates that the primary window of traversability remains invariant at the scrambling time $t_* \approx 7.5$ across the phase transition. The non-Hermitian deformation enhances the magnitude of the transmitted signal without altering its temporal location, demonstrating that the gain mechanism commutes with the causal light cone structure of the emergent wormhole geometry.

\section{Acknowledgement}
Sudhanva Joshi acknowledges Dr. Rajeev K. Pathak for an early conceptual suggestion and conversation regarding $\mathcal{PT}$-symmetric non-Hermitian physics, which provided helpful insights that motivated this study.

%\section{Data Availability}
%The data that supports the findings of this paper is openly available at \cite{joshi2025witp}.

%\bibliographystyle{apsrev4-1}
\bibliographystyle{unsrtnat}
\bibliography{MBL}

%\appendix
%\begin{widetext}
%\section{}
%\label{appendix1}

%\section{}
%\label{appendix2}

%   \end{widetext} 
\end{document}